# Study of using Quantum Computer to Solve Poisson Equation in Gate Insulators

Hector Jose Morrell
*Electrical Engineering*
*San Jose State University*
San Jose, USA
hector.morrell@sjsu.edu

Hiu Yung Wong\*
*Electrical Engineering*
*San Jose State University*
San Jose, USA
hiuyung.wong@sjsu.edu

*Abstract*—In this paper, the application of quantum computing (QC) in solving gate insulator Poisson equation is studied, through QC simulator and hardware in IBM. Various gate insulator stacks with and without fixed charges are studied. It is found that by increasing the number of clock bits and by choosing appropriate evolution time, accurate solutions can be obtained in QC simulation. However, when the real quantum hardware is used, the accuracy is substantially reduced. Therefore, a more robust quantum circuit or error correction should be employed and developed.

## I. INTRODUCTION

Quantum computing (QC) is becoming more promising and quantum supremacy has been demonstrated in a 53-qubit QC system [1]. One of the promising applications of QC is to speed up the solving of the system of linear equations, $A\vec{x} = \vec{b}$, in which vector $\vec{x}$ is solved for a given symmetric matrix, $A$, and a vector, $\vec{b}$. HHL algorithm [2][3], which has a time complexity of $O(\log N)$, is a QC algorithm that can provide exponential speedup over the classical conjugate gradient method. It is expected to have a big impact on various areas such as machine learning [4] and modeling of quantum systems [5]. It has also been proposed to solve the Poisson equation [6].

However, there is a lack of study of using QC to solve the Poisson equation in semiconductor problems. In this work, we study the performance of QC in solving the Poisson equation in gate dielectric stacks using QC simulator [7] and QC hardware in IBM [8]. $SiO_2$, $SiO_2/HfO_2$, and $Si_3N_4/SiO_2/Si_3N_4$ gate stacks with and without fixed charges are studied. Firstly, a system with an exact solution is studied in detail and solved using two different circuits in both QC simulation and hardware. Then the robustness of solving more complex systems is studied through QC simulation and the results are verified using TCAD Sentaurus [9].

## II. SIMULATION STRUCTURES

Poisson equations across various 1-D gate stacks are studied. Fig. 1 shows the structures simulated. Three types of structures, namely structure (a) ($Si_3N_4/SiO_2/Si_3N_4$), structure (b) ($SiO_2$), and structures (c1), (c2), (c3) ($SiO_2/HfO_2$), are studied. Fixed charges are put in structures (c2) and (c3). All structures are biased at 2V and are 2nm thick. The Poisson equation is discretized and the size of $A$ is 2×2 for structures (a), (b), and (c1), 4×4 for structure (c2), and 8×8 for structure (c3). Therefore, they can be handled by 1, 2, and 3 qubits respectively in quantum computing. Note that the terminal mesh points are at a fixed bias (Dirichlet boundary condition) and thus need not be solved. The equations are solved in Python 3.7.3 using direct solver, TCAD Sentaurus using Newton iteration, and Qiskit for quantum computing circuit simulation. Structure (a), which gives an exact solution is also implemented in IBM 5-qubit quantum computing hardware.

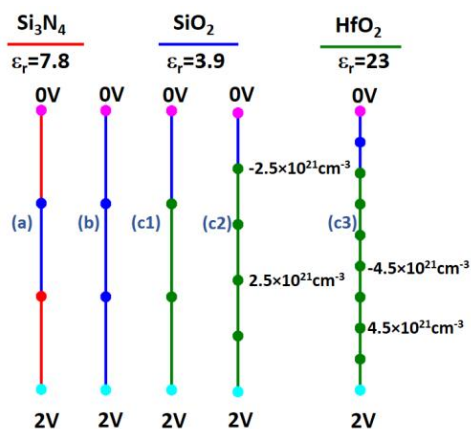

Figure 1: Gate stacks studied in this paper. Mesh points are showed in dots. Color legend of each material is shown on top with its relative dielectric constant. Fixed charge is also added to certain nodes.

## III. HHL ALGORITHM AND SIMULATION

### A. Overview

Fig. 2 shows the HHL algorithm implemented to solve structure (a). HHL Algorithm has 5 main sections: state preparation, quantum phase estimation (QPE), ancilla bit rotation, inverse quantum phase estimation, and measurement. In HHL, $A\vec{x} = \vec{b}$ is represented by $A|x\rangle = |b\rangle$. If $|u_i\rangle$ are the eigenvectors of $A$, the coefficient, $b_i$, in $\vec{b}$ is encoded as the coefficient of $|b\rangle$ in the basis formed by $|u_i\rangle$, i.e. $|b\rangle = \sum b_i |u_i\rangle$. The number of clock qubits ($n_l$) needs to be large enough to encode the eigenvalues of $A$. The algorithm starts with QPE by applying $e^{iAt}$ on $|b\rangle$ in order to find the eigenvalues, $\lambda_i$, of $A$, which are encoded in the clock qubits and approximated by $|\tilde{\lambda}_i\rangle$. $t$ is the evolution time whose effect will be studied in detail. After that, controlled rotation on the ancilla qubit by the clock qubits is performed and is followed by inverse QPE and the ancilla bit is measured. If the ancilla bit is measured to be $|0\rangle$, the result is discarded and the computation is repeated until the ancilla bit is measured to be $|1\rangle$ and the solution is encoded as the coefficients in the output qubit $|x\rangle$. One can prove that the solution of the system of linear equation is

$$x = \sum_{j=0}^{N-1} \frac{1}{\lambda_j} b_j |u_j\rangle \quad (1)$$

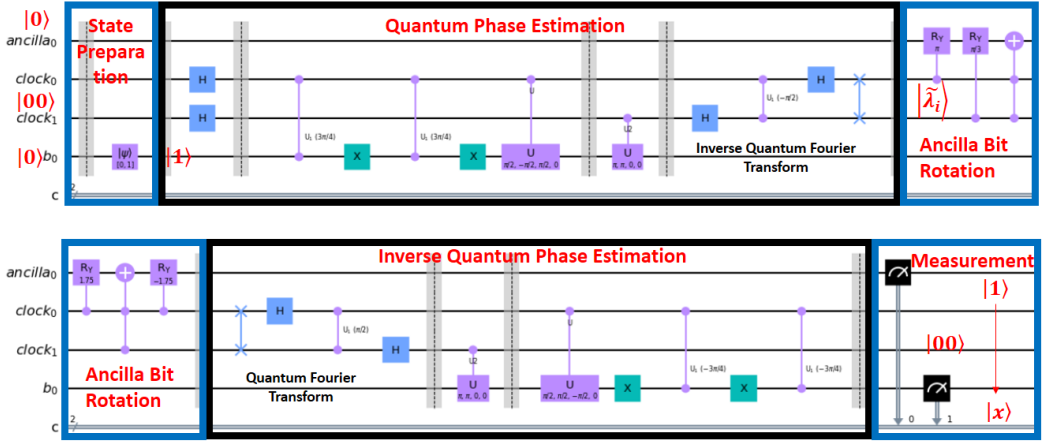

Figure 2: Quantum circuit used to solve structure (a). Boxes highlights various functional blocks.

*B. Example and Details*

As an example, for structure (a), after normalization, $\vec{b} = \begin{pmatrix} 0 \\ 1 \end{pmatrix}$, $A = \begin{pmatrix} 1 & -1/3 \\ -1/3 & 1 \end{pmatrix}$ with eigenvalues, $\lambda_1$ and $\lambda_2$, being 2/3 and 4/3, respectively.

$\vec{b}$ can be represented by 1 qubit, $|b\rangle$, and is obtained after applying X gate to $|0\rangle$. After the state is initialized, the eigenvalues are computed and encoded in the clock qubits using QPE which has 3 parts: superposition of the clock qubits (through Hadamard gates), controlled unitary operation, and Inverse Quantum Fourier Transform (IQFT). The clock qubits are in a superposition state, which in turn are controls for the unitary matrix applied to $|b\rangle$. The value of $n_l$ (number of clock qubits) is based on the desired resolution of the result. In this particular case, the eigenvalues can be encoded exactly using 2 clock qubits ($n_l = 2$) as $\lambda_1$ can be encoded as $|01\rangle$ and $\lambda_2$ can be encoded as $|10\rangle$ with a ratio of 2 unchanged.

The controlled unitary operations applied to $|b\rangle$ are $U = (e^{iAt})^{2^r}$. There are two controlled operations, each with $r = 0$ and 1, respectively. We use U-gate to perform the equivalent rotations,

$$U(\theta, \Phi, \lambda) = \begin{pmatrix} \cos\left(\frac{\theta}{2}\right) & -e^{i\lambda}\sin\left(\frac{\theta}{2}\right) \\ e^{i\Phi}\sin\left(\frac{\theta}{2}\right) & e^{i(\lambda+\Phi)}\cos\left(\frac{\theta}{2}\right) \end{pmatrix} \quad (2)$$

by setting $(\theta, \Phi, \lambda) = \left(\frac{\pi}{2}, -\frac{\pi}{2}, \frac{\pi}{2}\right)$ and $(\theta, \Phi, \lambda) = (\pi, \pi, 0)$ for $r = 0$ and 1, respectively. Appropriate phases are added before the operations.

We follow [10] and set $t=3/4\pi$ so that $\tilde{\lambda}_i = \lambda_i$ and an exact solution can be obtained. The IQFT, which is necessary to complete the QPE, is a standard one and would not be further explained in detail.

Including the ancilla qubit, this system has 4 qubits. The ancilla bit is added so that after the controlled rotation on the ancilla qubit by the clock qubits, the system has the following state (from left to right are the qubits from bottom to top in Fig. 2):

$$\sum_{j=0}^{N-1} b_j |u_j\rangle |\lambda_j\rangle \left(\sqrt{1 - \frac{C^2}{\lambda_j^2}}|0\rangle + \frac{C}{\lambda_j}|1\rangle\right) \quad (3)$$

And after applying inverse QPE, it becomes

$$\sum_{j=0}^{N-1} b_j |u_j\rangle |00\rangle \left(\sqrt{1 - \frac{C^2}{\lambda_j^2}}|0\rangle + \frac{C}{\lambda_j}|1\rangle\right) \quad (4)$$

If a measurement is performed on the ancilla bit and it returns $|1\rangle$ and if we only care about the 1st qubit (the bottom one in Fig. 2), the wavefunction collapses to

$$\sum_{j=0}^{N-1} \frac{C}{\lambda_j} b_j |u_j\rangle \quad (5)$$

Up to a normalization factor, this is the solution of the linear equation as shown in (1).

In order to achieve (3), controlled rotation $RY(\theta)$ is used to obtain $\frac{C}{\lambda_j}$,

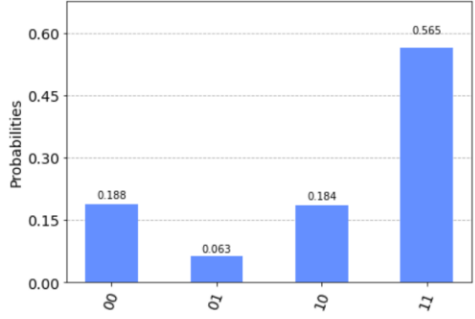

Figure 3: Simulation result of the circuit in Fig. 2. X-axis shows the values of the top most and bottom most qubits in Fig. 2. $|01\rangle$ to $|11\rangle$ ratio = 0.063:0.565=1:8.97.

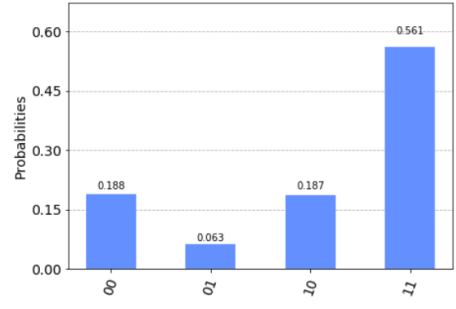

Figure 5: Simulated measurement probability of top most and bottom most bits in Fig. 4. $|01\rangle$ to $|11\rangle$ ratio = 0.063:0.561=1:8.9.

$$RY(\theta) = \exp\left(-i\frac{\theta}{2}Y\right) = \begin{pmatrix} \cos\left(\frac{\theta}{2}\right) & -\sin\left(\frac{\theta}{2}\right) \\ \sin\left(\frac{\theta}{2}\right) & \cos\left(\frac{\theta}{2}\right) \end{pmatrix} \quad (6)$$

The angle is found by using $\theta(\widetilde{c_1 c_0}) = 2\arcsin\left(\frac{C}{\widetilde{c_1 c_0}}\right)$, which is further approximated as $\theta(\widetilde{c_1 c_0}) = \pi c_0 + \frac{\pi}{3}c_1 - 3.51 c_0 c_1$, where $\widetilde{c_1 c_0}$ is the clock qubit in binary form for the C = 1 case. C = 1 is chosen because it maximizes the probability to obtain $|1\rangle$ when the ancilla bit is measured. Note that C must be larger than or equal to the smallest encoded eigenvalue, which is $|01\rangle$ = 1 in this case.

The circuit in Fig. 2 is simulated in software using "qasm_simulator". The solution of structure (a), i.e. the potential of the 2 interior points, are 0.5V and 1.5V, respectively. Therefore, the wavefunction amplitudes of $|x\rangle$ are expected to have a ratio of 1:3 and thus the probability of measuring $|x\rangle$ to be $|0\rangle$ or $|1\rangle$ has a ratio of 1:9. Fig. 3 shows that the ratio is 1:8.97 (Fig. 5) which is very close to the theoretical value.

It is possible to implement the same algorithm using different but equivalent circuits. We also implement the algorithm by modifying the circuit in [10]. The circuit is showed in Fig. 4. Fig. 5 shows the simulation results and the ratio between $|01\rangle$ to $|11\rangle$ is 1:8.9, which is also very close to the theoretical value and is similar to the result obtained in Fig. 3. Therefore, the circuits in Fig. 2 and Fig. 4 are considered to be equivalent.

## IV. HARDWARE IMPLEMENTATION

Structure (a) has a matrix that can give an exact solution in QC. Therefore, the quantum circuits in Fig. 2 and Fig. 4 are implemented to solve the Poisson equation in structure (a) and submitted to run in IBM Quantum Computers.

Three 5-qubit quantum computers have been tried, namely, *ibmq_5_yorktown*, *ibmq_belem,* and *ibmq_santiago*, which have quantum volumes of 8, 16, and 32 respectively. *ibmq_5_yorktown* has the best result and is shown in Fig. 6 and Fig. 7. However, the result is far from the expected value. The probability of measuring $|x\rangle$ to be $|0\rangle$ or $|1\rangle$ is found to be 1:1.41 and 1:1, respectively for circuits in Fig. 2 and Fig. 4, instead of 1:9. This is probably due to noise and the lack of error coding. Note that the results also have a substantial difference when it is run on different days.

## V. SOFTWARE SIMULATION ACCURACY

We further study the effect of $t$ and $n_l$ on the fidelity [11] and error using the built-in HHL function in [7] (python 3.7.3 and qiskit 0.23.5) using "statevector_simulator" on larger systems in Fig. 1. Fig. 8 shows that state fidelity is not a good metric to measure accuracy. One may achieve close to 100% fidelity but with an error (defined by average relative absolute error) close to 16% (structure (b)). It also shows that by increasing $n_l$, the error is reduced and it gives a larger window of $t$ to achieve a small error. Fig. 9 compares the QC solutions for structures (c2) and (c3) against the TCAD solution. For structure (c3), 10 qubits are needed but there is still a substantial error compared to TCAD as the eigenvalue cannot be encoded exactly in the 6 clock qubits.

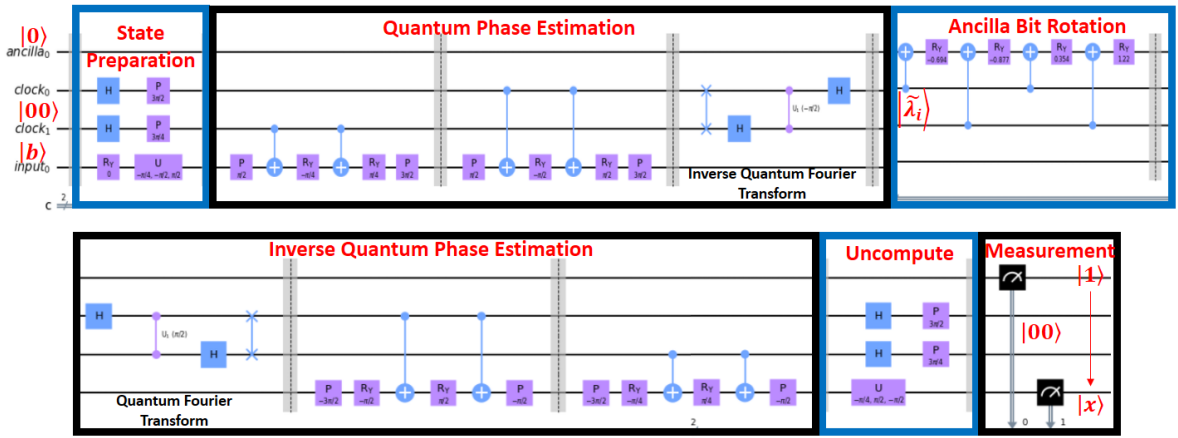

Figure 4: Quantum circuit modified from [10] used to solve structure (a). Boxes highlights various functional blocks.

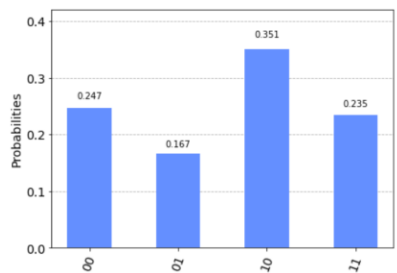

Figure 6: Hardware measurement probability of top most and bottom most bits in Fig. 2. $|01\rangle$ to $|11\rangle$ ratio = 0.167:0.235=1:1.41.

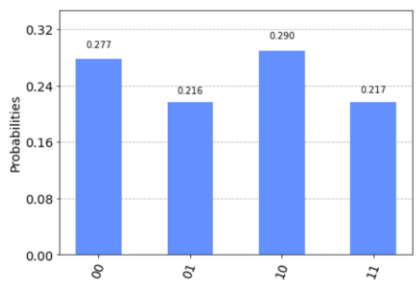

Figure 7: Hardware measurement probability of top most and bottom most bits in Fig. 4. $|01\rangle$ to $|11\rangle$ ratio = 0.216:0.217=1:1.

## VI. CONCLUSIONS

In this paper, we studied the use of the quantum computer through both simulation and hardware implementation to solve the Poisson equation in various gate stacks. We show that fidelity is not a good metric to gauge accuracy. While simulation shows that quantum computers can achieve similar accuracy as in TCAD, hardware implementation is not accurate enough, and probably more sophisticated error correction is needed.

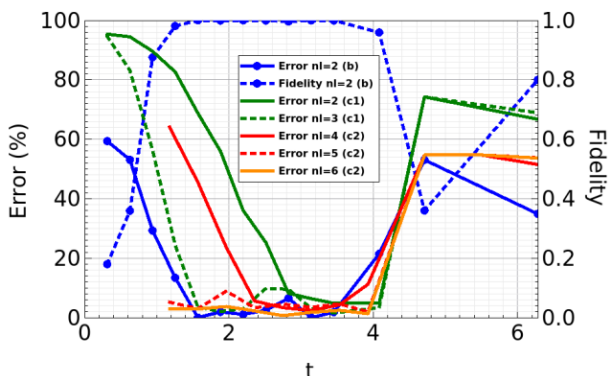

Figure 8: Simulated QC fidelity and error as a function of $t$ and $n_l$ of structures (b), (c1), and (c2). Error is defined by the average relative absolute error.

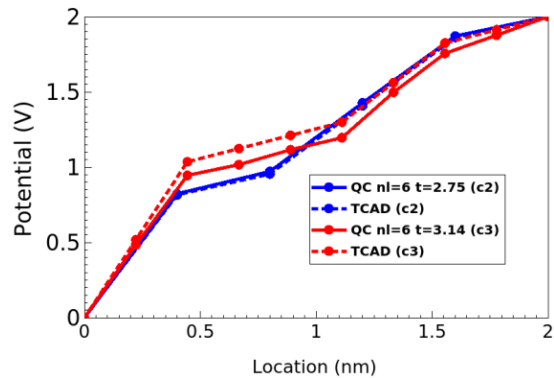

Figure 9: Best solutions obtained from QC simulation and TCAD for structures (c2) and (c3).

### ACKNOWLEDGMENT

The research work is benefited from the "SJSU-IBM Acceleration: Quantum Classrooms" project. This research is partially supported by the Level-Up Grant, Office of Research, San Jose State University. Some of the materials are based upon work supported by the National Science Foundation under Grant No. 2046220.